\documentclass[runningheads]{llncs}
\usepackage{amsfonts}
\usepackage{amssymb}
\usepackage{multirow}
\usepackage{graphicx}
\usepackage{cite} 
\usepackage{color}

\begin{document}
\title{SelfCoLearn: Self-supervised collaborative \\learning for accelerating dynamic MR imaging} 

\author{Juan Zou\inst{1,2}\and Cheng Li\inst{2}\and Sen Jia\inst{2} \and Ruoyou Wu\inst{2,3}\and Tingrui Pei\inst{1,4}\and Hairong Zheng\inst{2,3} \and Shanshan Wang\inst{2,3}}
\authorrunning{Juan Zou et al.}
\institute{Xiangtan University, Xiangtan Hunan, China \and Paul C. Lauterbur Research Center for Biomedical Imaging, Shenzhen Institutes of Advanced Technology, Chinese Academy of Sciences, Shenzhen, Guangdong, China
\and Pengcheng Laboratory, Shenzhen
, Guangdong, China
\and Jinan University, Guangzhou, Guangdong, China
\\
\email{peitr@163.com, Sophiasswang@hotmail.com, ss.wang@siat.ac.cn}}

\maketitle

\begin{abstract}
Lately, deep learning has been extensively investigated for accelerating dynamic magnetic resonance (MR) imaging, with encouraging progresses achieved. However, without fully sampled reference data for training, current approaches may have limited abilities in recovering fine details or structures. To address this challenge, this paper proposes a self-supervised collaborative learning framework (SelfCoLearn) for accurate dynamic MR image reconstruction from undersampled k-space data. The proposed framework is equipped with three important components, namely, dual-network collaborative learning, reunderampling data augmentation and a specially designed co-training loss. The framework is flexible to be integrated with both data-driven networks and model-based iterative un-rolled networks. Our method has been evaluated on in-vivo dataset and compared to four state-of-the-art methods. Results show that our method possesses strong capabilities in capturing essential and inherent representations for direct reconstructions from the undersampled k-space data and thus enables high-quality and fast dynamic MR imaging.
\end{abstract}

\section{Introduction}

Deep learning-based dynamic magnetic resonance (MR) imaging has attracted substantial attention in recent years. It draws knowledge from big datasets via network training and then uses the trained network to conduct dynamic image reconstruction from the undersampled k-space data. Compared to the classical compressed sensing methods \cite{gamper2008compressed,zhao2012image,jung2007improved,wang2013compressed,caballero2014dictionary,ji2008dynamic,zhao2019motion,jung2009k,otazo2015low,huang2005k}, deep learning-based methods have made encouraging performances and progresses. 

Based on the reliance on the fully sampled dataset or not, existing methods can be roughly categorized into two types, fully-supervised methods and unsupervised ones. For the fully-supervised methods, data pairs are needed for the training of the neural networks between the undersampled/corrupted data and the fully sampled/ reference data \cite{wang2016accelerating, zhang2018ista,eo2018kiki,sun2016deep,aggarwal2018modl,hammernik2018learning,zhu2018image,akccakaya2019scan,mardani2018deep,cohen2018mr,yoon2018quantitative}. In this category, different network structures and prior knowledge have been explored \cite{huang2021dynamic,seegoolam2019exploiting,qin2019k,schlemper2017deep,qin2018convolutional,qin2021complementary,huang2021deep,ke2021learned}. For example, Schlemper et al. \cite{schlemper2017deep} proposed a cascade network architecture composed of an intermediate de-aliasing convolutional neural network (CNN) module and a data consistency layer. Chen et al. \cite{qin2018convolutional} applied bidirectional convolutional recurrent neural network (CRNN) with interleaved data consistency to accelerate MR imaging. Chen et al. \cite{qin2021complementary} designed a parallel framework, including a time-frequency domain CRNN and an image domain CRNN to simultaneously exploit spatiotemporal correlations. Huang et al.\cite{huang2021deep} unrolled the low-rank plus sparse method \cite{otazo2015low} into a deep neural network to learn the low rank and sparse regularization. Ke et al. \cite{ke2021learned} exploited the low rank priors (SLR-Net). These methods have made great progresses in accelerating dynamic MR imaging. However, one major challenge of these methods is that, in many practical imaging scenarios, obtaining high-quality fully sampled dynamic MRI data is infeasible due to various factors, such as physiological motions of patients, imaging speed restriction, etc. Therefore, the requirement of fully sampled reference data for network training has hindered the wide application of these methods.

To address this issue, researchers have developed unsupervised learning methods to train models without fully sampled reference data \cite{ke2020unsupervised,yoo2021time,acar2021self}. For example, Ke et al. \cite{ke2020unsupervised} generated a pseudo reference label from undersampled data by merging neighbouring frames of undersampled k-space data. Jin et al. \cite{yoo2021time} extended the framework of deep image prior \cite{ulyanov2018deep} to dynamic non-cartesian MRI. Recently, Yaman et al. \cite{yaman2020self} proposed a classical self-supervised learning method (SSDU) for static MR imaging. SSDU divides the acquired undersampled data into two parts. One part is treated as the input data, and the other is utilized as the supervisory signals \cite{akccakaya2022unsupervised}. Subsequently, Acar et al. \cite{acar2021self} applied SSDU to reconstruct dynamic MR images. All of these works have made great contributions to unsupervised dynamic MR image reconstruction. Nevertheless, these works still have spaces to improve in recovering fine details or structures due to the incomplete inherent representation of the undersampled data compared to fully sampled data. 

To boost the performances for accelerating dynamic MR imaging without fully sampled reference data, we propose a self-supervised collaborative learning framework (SelfCoLearn). Taking the assumption that the latent representation of network predictions is consistent under different reundersampling data augmentation from the same data. SelfCoLearn apply collaborative training of dual-network with reundersampling data augmentation to explore more sufficient prior knowledge compared to a single network. Specifically, from undersampled k-space data, the operations of reundersampling data augmentation are implemented to obtain two reundersampling inputs for dual-network. Dual networks are trained collaboratively with a specially designed co-training loss in an end-to-end manner. With this collaborative training strategy, the proposed framework can possess strong capabilities in capturing essential and inherent representations from the undersamled k-space data in self-supervised learning manner. Additionally, the framework is flexible to be integrated with both data-driven networks and model-based iterative un-rolled networks \cite{8962949} for dynamic MR imaging. The main contributions can be described as follows:

\begin{itemize}
    \item We present a self-supervised collaborative learning framework with reundersampling data augmentation for accelerating dynamic MR imaging. The proposed framework is flexible to be integrated with both data-driven networks and iterative un-rolled networks.
    \item A co-training loss, which includes an undersampled consistency loss term and a contrastive consistency loss term, is designed to guide the end-to-end framework to capture essential and inherent representations from the undersamled k-space data.
    \item Extensive experiments are conducted to evaluate the effectiveness of the proposed SelfCoLearn with both data-driven and iterative un-rolled networks, with more promising results obtained compared to four state-of-the-art methods.
\end{itemize}

The remainder of this paper is organized as follows: Section II states the dynamic MR imaging problem and the proposed SelfCoLearn with different backbone networks. Section III summarizes the experiments and results to demonstrate the effectiveness of the proposed SelfCoLearn, while discussion is presented in section IV. Section V concludes the work.

\section{Methodology}
\subsection{Dynamic MR Imaging Formulation}
The goal of dynamic MR imaging is to estimate dynamic MR image sequences $\mathbf{x} \in \mathbb{C}^{N}$ from undersampled measurements $\mathbf{y} \in \mathbb{C}^{M}(M \ll N)$ in k-space. $N=N_{h}N_{W}T$ is a vector. $N_{h}$ and $N_{W}$ are  height and width of the frame respectively. $T$ represents the number of frames in each sequence. Thus, the imaging model can be described as follows:
\begin{equation}
\label{deqn_ex1}
\mathrm{y}=\mathbf{A} \mathbf{x}+\mathrm{e}
\end{equation}
where $\mathrm{e} \in \mathbb{C}^{M}$ is noise and $\mathbf{A}=\mathbf{P F}$ is an undersampled Fourier encoding operator, $\mathbf{F}$ is 2D Fourier transform to each frame in dynamic image sequence and $\mathbf{P}$ is the undersampled mask for each frame. In general, the reconstruction problem is formulated as an unconstrained optimization problem to explore the prior knowledge:
\begin{equation}
\label{deqn_ex2}
\mathrm{x}^{*}=\arg \min _{\mathrm{x}}\frac{1}{2}\|\mathbf{A} \mathbf{x}-\mathrm{y}\|_{2}^{2}+\lambda \mathcal{R}(\mathbf{x})
\end{equation}
where $\mathcal{R}(\mathbf{x})$ represents a prior regularization item on $\mathbf{x}$, and $\lambda$ is the weight of the regularization. $\frac{1}{2}\|\mathbf{A} \mathbf{x}-\mathrm{y}\|_{2}^{2}$ is the data fidelity item, which ensures the reconstruction result to be consistent with the original undersampled measurements.

For fully-supervised deep learning methods, it typically uses a CNN $f_{C N N}\left(\mathbf{y} \mid \theta\right)$, as a regularization term $\mathcal{R}(\mathbf{x})$, by learning the mapping between undersampled/ corrupted data and their corresponding fully sampled data with parameters $\theta$. Its mathematical description can be given as:

\begin{equation}
\label{deqn_ex3}
\arg\min _{\theta} \sum_{i=1}^{S}  \mathcal{L}\left(f_{C N N}\left(\mathbf{y}_{i} \mid \theta\right), \mathbf{x}_{i}^{r e f}\right)
\end{equation}
where $S$ is the number of dynamic image sequences in the training set. $\mathbf{x}_{i}^{r e f}$ is the fully sampled data of the subject data $i$. $\mathcal{L}(\cdot)$ denotes the loss function between the predicted reconstruction output and the fully sampled reference data, which typically adopts the $l_{1}-$norm or $l_{2}-$norm.

\subsection{The Overall Framework}
\begin{figure*}[htbp] 
\centering
\includegraphics[width=\textwidth]{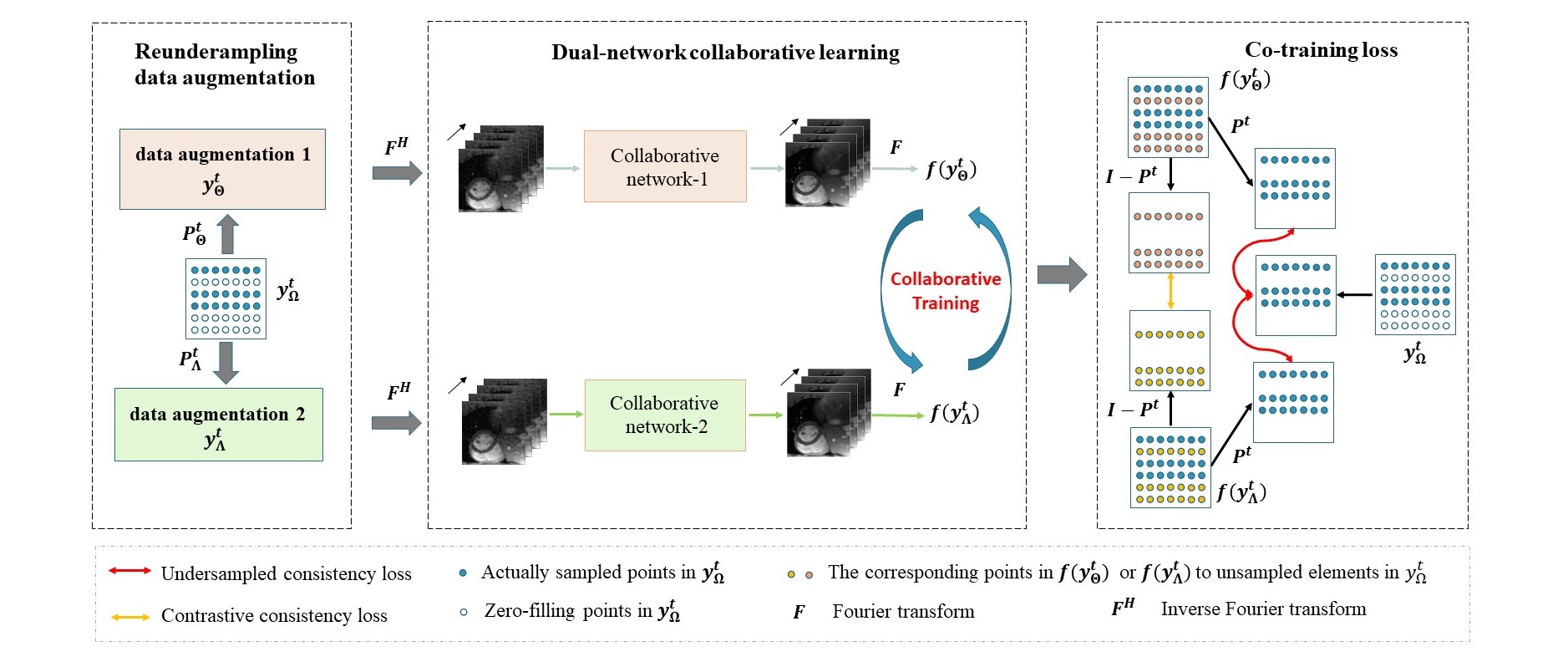} 
\caption{An overview of the proposed self-supervised collaborative training framework. The raw undersampled k-space data sequence $\mathbf{y}_{\Omega}^{t}$ is undersampled from the fully sampled data with the undersampled mask $P^{t}$. K-space data sequence $\mathbf{y}_{\Theta}^{t}$ and $\mathbf{y}_{\Lambda}^{t}$ are reundersampled from $\mathbf{y}_{\Omega}^{t}$ with reundersampled mask $\mathbf{P}_{\Theta}^{t}$ and $\mathbf{P}_{\Lambda}^{t}$. The two networks received inputs from zero-filling image sequences of $\mathbf{y}_{\Theta}^{t}$ and $\mathbf{y}_{\Lambda}^{t}$, respectively. The predicted image sequences of networks are transformed to k-space data $f\left(y_{\Theta}^{t}\right)$ and $f\left(y_{\Lambda}^{t}\right)$ with two-dimensional Fourier transform. A co-training loss is calculated utilizing $\mathbf{y}_{\Omega}^{t}$, $f\left(y_{\Theta}^{t}\right)$ and $f\left(y_{\Lambda}^{t}\right)$. The backbone network can flexibly adopt both data-driven network or iterative un-rolled network, such as CRNN, k-t NEXT and SLR-Net et al. Collaborative network-1 and collaborative network-2 have the same network structure but with different weight parameters ${\theta}_{\Theta}$ and ${\theta}_{\Lambda}$ respectively.}
\label{fig1}
\end{figure*}

In our work, we propose a simple but effective self-supervised training framework for dynamic MR imaging, whose paradigm is shown in Fig. \ref{fig1}. Our framework trains two independent reconstruction networks simultaneously, which have different inputs and different weight parameters. The backbone network can adopt either data-driven network architecture or the iterative un-rolled network, such as CRNN \cite{qin2018convolutional}, k-t NEXT \cite{qin2019k} and SLR-Net \cite{ke2021learned} et al. Based on the consistency between two network predictions, the network provides complementary information for the to-be-reconstructed dynamic MR images in its peer partner, which is an additional regularization compared to the existing unsupervised methods \cite{acar2021self}. The two networks will finally realize consistent reconstruction in the training process. Specifically, given a raw undersampled k-space data sequence $\Omega=\left\{\mathbf{y}_{\Omega}^{t}\right\}_{t=1}^{T}$, we reundersampled the original k-space data $\mathbf{y}_{\Omega}^{t}$ to construct a partial data points sequence $\left\{\mathbf{y}_{\mathrm{u}}^{t}\right\}_{t=1}^{T}$:
\begin{equation}
\label{deqn_ex4}
\mathbf{y}_{\mathrm{u}}^{t}=P_{\mathrm{u}}^{t}\left(\mathbf{y}_{\Omega}^{t}\right), \mathrm{t}=1, \ldots, \mathrm{T}, \mathrm{u}=\Theta, \Lambda
\end{equation}
where $t$ is the index of sequence, $u$ denotes the index of the two training sequences and $P_{\mathrm{u}}^{t}$ is the undersampled mask for frame $t$. In order to make full use of all data points in $\mathbf{y}_{\Omega}^{t}$ to learn representation, and ensure each network can provide complementary information for the to-be-reconstructed dynamic MR images in its peer network, these training sequences are generated adhere to the following data augmented principles: (1) The union of data points in two training sequences must be equal to the data $\mathbf{y}_{\Omega}^{t}$, i.e., $\mathbf{y}_{\Omega}^{t}=\mathbf{y}_{\Theta}^{t} \cup \mathbf{y}_{\Lambda}^{t}$. (2) The data points in two training sequences should be different, i.e., $\mathbf{y}_{\Theta}^{t} \neq \mathbf{y}_{\Lambda}^{t}$. (3) The training sequences should include most of the low frequency data points and part of the high frequency data points. Following these principles, the two training sequences contain similar data points in the low frequency region, and different points in the high frequency region. Noted that the operation of data reundersampling is only necessary during training, the reconstructed images can be inferred from the test data directly.

\subsection{Network Architectures}
{\itshape 1) Data-driven  dynamic MR imaging:} In the data-driven settings, the common practice is to decouple Eq. \ref{deqn_ex2} into a regularization term and a data fidelity term via utilizing the variable splitting technique \cite{schlemper2017deep,qin2018convolutional}. By introducing an auxiliary variable $\mathbf{z}=\mathbf{x}$, Eq.\ref{deqn_ex2} can be re-formulated as the penalty function:
\begin{equation}
\label{deqn_ex5}
\arg \min _{\mathrm{x,z}} \lambda\mathcal{R}(\mathbf{z})+\frac{1}{2}\|\mathbf{A} \mathbf{x}-\mathrm{y}\|_{2}^{2}+\mu\|\mathbf{x}-\mathbf{z}\|_{2}^{2}
\end{equation}
where $\mu$ is a penalty parameter. Eq.\ref{deqn_ex5} can then be solved iteratively via alternating minimization over $\mathbf{z}$ and $\mathbf{x}$:

\begin{equation}
\label{deqn_ex6a}
\mathbf{z}^{n}=\arg \min _{\mathrm{z}} \lambda\mathcal{R}(\mathbf{z})+\mu\|\mathbf{x}^{n-1}-\mathbf{z}\|_{2}^{2}
\end{equation}
\begin{equation}
\label{deqn_ex6b}
\mathbf{x}^{n}=\arg \min _{\mathrm{x}} \frac{1}{2}\|\mathbf{A} \mathbf{x}-\mathrm{y}\|_{2}^{2}+\mu\|\mathbf{x}-\mathbf{z}^{n}\|_{2}^{2}
\end{equation}
where $n \in\left\{0,1,2, \ldots, N-1\right\}$ is the $n$th iteration, $\mathbf{x}^{0}$ is the zero-filling image transformed from original undersampled measurement, $\mathbf{z}^{n}$ denotes the intermediate reconstruction sequence, and $\mathbf{x}^{n}$ denotes the final reconstruction sequence at each iteration. In Eq.\ref{deqn_ex6b}, the operation on the intermediate reconstruction sequence $\mathbf{z}^{n}$ is a data consistency step, which uses the original sampled k-space data points to replace the corresponding data points in the reconstructed k-space data \cite{schlemper2017deep}. The iterative optimization process in Eq.\ref{deqn_ex6a} and Eq.\ref{deqn_ex6b} is unrolled into a neural network. 

\begin{figure}[htbp] 
\centering\includegraphics[width=0.5\textwidth]{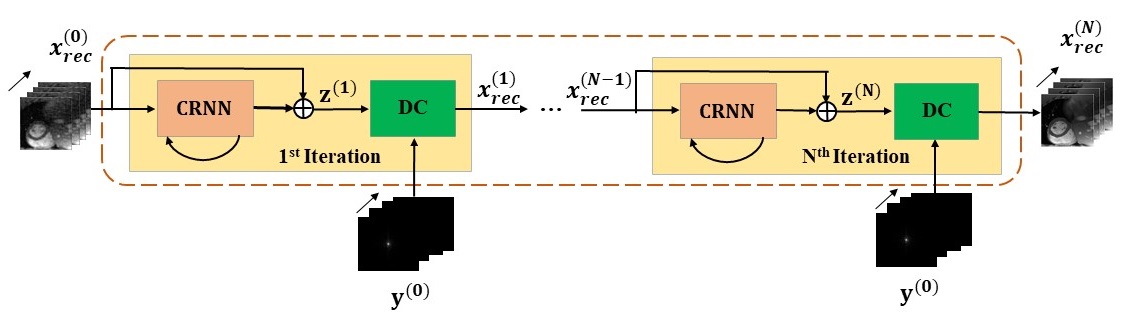} 
\caption{The unrolled architecture of CRNN with N iterations.}
\label{fig2}
\end{figure}

CRNN \cite{qin2018convolutional} is a typical data-driven method that integrates data consistency in k-space. A single iteration of CRNN can be illustrated as the following process:
\begin{equation}
\label{deqn_ex7a}
\mathbf{x}_{r n n}^{(n)}=\mathbf{x}_{r e c}^{(n-1)}+\mathrm{CRNN}\left(\mathbf{x}_{r e c}^{(n-1)}\right)
\end{equation}
\begin{equation}
\label{deqn_ex7b}
\mathbf{x}_{r e c}^{(n)}=\mathrm{DC}\left(\mathbf{x}_{r n n}^{(n)} ; \mathrm{y}, \lambda\right)
\end{equation}
where $\mathbf{x}_{r n n}^{(n)}$ is the intermediate reconstruction sequence analogous to $\mathbf{z}^{n}$ in Eq.\ref{deqn_ex6a}, and $\mathbf{x}_{r e c}^{(n)}$ denotes the final predicted result at each iteration analogous to $\mathbf{x}^{n}$ in Eq.\ref{deqn_ex6b}. The regularization subproblem in Eq.\ref{deqn_ex6a} is solved by using a convolutional recurrent neural network. The data consistency subproblem in Eq.\ref{deqn_ex6b} is treated as a data consistency network layer. The unrolled architecture of CRNN is shown in Fig. \ref{fig2}. More details of CRNN layers can be found in \cite{qin2018convolutional}.

{\itshape 2) Un-rolled dynamic MR imaging:} Another widely-used strategy is un-rolled dynamic MR imaging, which constructs CNNs according to the iterations of traditional optimization algorithms. Different optimization algorithms lead to different network architectures. SLR-Net, which formulates sparse and low rank priors as regularized terms in an optimization algorithm \cite{ke2021learned}, is a typical example of un-rolled method. In SLR-Net, by introducing an auxiliary variable $\mathbf{M}$, Eq.\ref{deqn_ex2} can be decoupled as the fidelity term, sparse regularization term and the low rank regularization term:

\begin{equation}
\label{deqn_ex8}
\arg \min _{\mathrm{x,M}}\frac{1}{2}\|\mathbf{A} \mathbf{x}-\mathrm{y}\|_{2}^{2}+\lambda_{1}\|D \mathbf{x}\|_{1}+\lambda_{2}\|M \|_{*}
\end{equation}
where $D$ is a sparse transform in a certain sparse domain. $M=R\mathbf{x}$ is a matrix (with size ($N_\mathbf{h} \times N_\mathbf{w}$, $T$)), in which each column corresponds to one frame in dynamic MR image sequence. $R$ is a reshaping operator. $ \|M \|_{*}$ is the nuclear norm. Previous works have proven that nuclear norm minimization is effective in low-rank matrix recovery \cite{candes2009exact}. More details of the iterative process in SLR-Net can be found in \cite{ke2021learned}.

\subsection{The Proposed Co-training Loss}
We have designed a co-training loss to promote accurate dynamic MR image reconstruction in a self-supervised manner. The core idea of the co-training loss is to enforce the consistency not only between the reconstruction results and the original undersampled k-space data, but also between two network predictions. Compared with existing methods with single network, the consistency between two network predictions is an additional regularization, which guides the dual-network to learn more correct information. Specifically, the co-training loss in SelfCoLearn, including an undersampled consistency loss term and a contrastive consistency loss term, is calculated to optimize the whole framework.

Let $f_{SelfCoLearn}\left(\mathbf{y}_{\Omega}^{t}\right)$ denotes SelfCoLearn, $\mathbf{y}_{\Omega}^{t}$ is the original undersampled k-space data. During training, two training sequences $\mathbf{y}_{\Theta}^{t}$ and $\mathbf{y}_{\Lambda}^{t}$ are generated from $\mathbf{y}_{\Omega}^{t}$ following the data augmented principles in section II-B as follows:
\begin{equation}
\label{deqn_ex11}
\mathbf{y}_{\Theta}^{t}=\mathbf{P}_{\Theta}^{t} \mathbf{y}_{\Omega}^{t},\quad \mathbf{y}_{\Lambda}^{t}=\mathbf{P}_{\Lambda}^{t} \mathbf{y}_{\Omega}^{t},
\end{equation}
where $\mathbf{P}_{\Theta}^{t}$ and $\mathbf{P}_{\Lambda}^{t}$ are reundersampled mask for $\mathbf{y}_{\Omega}^{t}$. The undersampled consistency loss is mainly referred to the actually sampled k-space points in $\mathbf{y}_{\Omega}^{t}$, which ensures the corresponding sampled points in network prediction consistent with actually sampled k-space points in $\mathbf{y}_{\Omega}^{t}$. The corresponding sampled points $\mathrm{y}_{\Theta \rightarrow \Omega}^{t}$ and $\mathrm{y}_{\Lambda \rightarrow \Omega}^{t}$ in these two network predictions can be written as:
\begin{equation}
\label{deqn_ex12}
\mathrm{y}_{\Theta \rightarrow \Omega}^{t}=\mathbf{P}^{t} f \left( \mathbf{y}_{\Theta}^{t} \right),\quad \mathrm{y}_{\Lambda \rightarrow \Omega}^{t}= \mathbf{P}^{t} f \left( \mathbf{y}_{\Lambda}^{t} \right),
\end{equation}
where k-space data $f \left( \mathbf{y}_{\Theta}^{t} \right)$ and $f \left( \mathbf{y}_{\Lambda}^{t} \right)$ are transformed from the predicted image sequences of two networks, respectively. $\mathbf{P}^{t}$ is the undersampled mask, which is applied to generate the original undersampled k-space data $\mathbf{y}_{\Omega}^{t}$ from the fully sampled data. Thus, the Undersampled Consistency loss term is used to calculate the mean-square-error between the actually sampled k-space points in $\mathbf{y}_{\Omega}^{t}$ and that in each network prediction as follows:
\begin{equation}
\label{deqn_ex13}
\mathcal{L}_{U C}=\left\|\mathbf{y}_{\Theta \rightarrow \Omega}^{t}-\mathbf{y}_{\Omega}^{t}\right\|_{2}^{2}+\left\|\mathbf{y}_{\Lambda \rightarrow \Omega}^{t}-\mathbf{y}_{\Omega}^{t}\right\|_{2}^{2}.
\end{equation}

In the ideal case, when different reundersampled k-space data from the same data are feed into two networks, the network predictions shall approximate the fully sampled reference data with the network optimizations. However, when the fully sampled reference data are absent, these two networks may produce different predicted results only with the undersampled consistency loss, and result in different reconstruction performances. As mentioned above, a contrastive consistency loss is defined to compute the mean-square-error between two network predictions with different reundersampling inputs from the same data. Specially, our contrastive consistency loss term is mainly referred to the corresponding points in network predictions to unsampled k-space points in $\mathbf{y}_{\Omega}^{t}$. These corresponding points $\bar{\mathbf{y}}_{\Theta \rightarrow \Omega}^{t}$ and $\bar{\mathbf{y}}_{\Lambda \rightarrow \Omega}^{t}$ in two network predictions $f \left( \mathbf{y}_{\Theta}^{t} \right)$ and $f \left( \mathbf{y}_{\Lambda}^{t} \right)$ can be written as: 
\begin{equation}
\label{deqn_ex14}
\bar{\mathbf{y}}_{\Theta \rightarrow \Omega}^{t}=\left(\mathbf{I}-\mathbf{P}^{t}\right) f \left( \mathbf{y}_{\Theta}^{t} \right), \quad \bar{\mathbf{y}}_{\Lambda \rightarrow \Omega}^{t}=\left(\mathbf{I}-\mathbf{P}^{t}\right) f \left( \mathbf{y}_{\Lambda}^{t} \right),
\end{equation}
therefore, the Contrastive Consistency loss term is formulated as:
\begin{equation}
\label{deqn_ex15}
\mathcal{L}_{C C}=\left\|\bar{\mathrm{y}}_{\Theta \rightarrow \Omega}^{t}-\bar{\mathrm{y}}_{\Lambda \rightarrow \Omega}^{t}\right\|_{2}^{2}.
\end{equation}
Combining the two loss terms, our final co-training loss function can be written as:
\begin{equation}
\label{deqn_ex16}
\mathcal{L}_{c o}=\mathcal{L}_{U C}+\gamma \mathcal{L}_{C C},
\end{equation}
where $\gamma$ is used to balance the weight parameter of undersampled consistency loss and contrastive consistency loss. During testing phase, undersampled data sequence is set as input of collaborative network-1 to obtain the final reconstruction result.

\section{Experimental Results}

Extensive experiments have been performed to evaluate the effectiveness of the proposed method. The performance of SelfCoLearn is compared with that of four state-of-the-art fully-supervised and self-supervised learning methods at different acceleration factors. Besides, SelfCoLearn with different backbone networks have been experimented, including both data-driven networks and iterative un-rolled networks for dynamic MR imaging. Then, results of ablation studies are reported to investigate the impacts of the undersampled consistency loss term and contrastive consistency loss term. Finally, reconstruction results with different co-training loss calculated in different domains are reported to further validate the effectiveness of proposed SelfCoLearn.
\subsection{Experimental Setup}
{\itshape 1) Dataset:} T1-weighted FLASH sequence is utilized to collect fully sampled cardiac data from 101 volunteers on a 3T scanner. All in vivo experiments have been approved by the Institutional Review Board (IRB) of Shenzhen Institutes of Advanced Technology, and written informed consent have been obtained from all volunteers. Each scan acquires a single slice from the volunteer with 25 temporal frames. The following parameters were used for the FLASH sequence: FOV 330×330 mm, acquisition matrix 192×192, slice thickness = 6 mm, TR/TE = 50 ms/3 ms, and 24 receiving coils. The raw multi-coil data of each frame was combined by the adaptive coil combine method \cite{lee2010admira} to produce a single-channel complex-valued reconstruction image. Then, the complex-valued images were transformed to k-space data, which simulate a fully sampled single-coil data acquisition. Training of neural networks requires a large amount of data. To this end, we use data augmentation to enlarge the data set. Specifically, we translated the original images along x, y, and t directions, and the translation step size is 128×128×14, and the stride along the three directions are 12, 12, and 3, respectively. Finally, our dataset consists of 6214 complex-valued cardiac MR data sequences with size 128×128×14. 5950 cardiac MR data sequences were randomly selected as the training dataset, 50 cardiac sequences were used as validation dataset, and the remaining sequences were used for testing.

{\itshape 2) Reunderampling K-space Data Augmentation:} In the proposed method, the fully sampled data are only used to generate the original undersampled k-space data $\mathbf{y}_{\Omega}^{t}$ with a 2D random retrospective undersampled mask $\mathbf{P}^{t}$. Following the principles of training data augmentation in section II-B, $\mathbf{y}_{\Omega}^{t}$ is augmented to two training sequences $\mathbf{y}_{\Theta}^{t}$ and $\mathbf{y}_{\Lambda}^{t}$ with two 2D random reundersampled masks $\mathbf{P}_{\Theta}^{t}$ and $\mathbf{P}_{\Lambda}^{t}$. $\mathbf{P}_{\Theta}^{t}$ with 2-fold acceleration is used for collaborative network-1, and $\mathbf{P}_{\Lambda}^{t}$, which is combines the complementary set of $\mathbf{P}_{\Theta}^{t}$ with some low-frequency data points of $\mathbf{P}^{t}$, is used for collaborative network-2.

{\itshape 3) Evaluation Metrics:} To evaluate the reconstruction performance, the mean-square-error (MSE), peak-signal-to-noise ratio (PSNR), and structural similarity index (SSIM) \cite{wang2004image} are calculated as follows:
\begin{equation}
\label{deqn_ex17}
\mathrm{M S E}=\|\mathrm{Ref} -\mathrm{Rec}\|_{2}^{2}
\end{equation}
\begin{equation}
\label{deqn_ex18}
\mathrm{P S N R}=20 \log _{10} \frac{MAX_{R e f}}{\sqrt{M S E}}
\end{equation}
\begin{equation}
\label{deqn_ex19}
\mathrm{S S I M}=\frac{\left(2 \mu_{R e f} \mu_{R e c}+c_{1}\right)\left(2 \sigma_{R e f, R e c}+c_{2}\right)}{\left(\mu_{R e f}^{2}+\mu_{R e c}^{2}+c_{1}\right)\left(\sigma_{R e f}^{2}+\sigma_{R e c}^{2}+c_{2}\right)}
\end{equation}
where $\mathrm{Rec}$ is the reconstructed image sequence, and $\mathrm{Ref}$ represents the reference image sequence. $\mathrm{MAX}_{R e f}$ is the maximum possible value in the image. $\mu_{R e f}$ and $\mu_{R e c}$ are the averaged intensity values of the corresponding images. $\sigma_{R e f}$ and $\sigma_{R e c}$ are the variances. $c_{1}$ and $c_{2}$ are adjustable constants. $\sigma_{R e f, R e c}$ is the covariance. The SSIM index is a multiplicative combination of the luminance term, the contrast term, and the structural term (details shown in \cite{wang2004image}).

{\itshape 4) Model Configuration and Implementation Details:}  The proposed framework is flexible to be integrated with both data-driven and iterative un-rolled networks. Most of our experiments adopt CRNN as the backbone network. In detail, the network is composed of a bidirectional CRNN layer, three CRNN layers, a 2D CNN layer, a residual connection sums output with input and a DC layer. The nonlinear activation function utilized is the Rectified linear unit (ReLU). For the bidirectional CRNN and CRNN layer, the number of convolution filters is set to $n_{f}=64$ with a kernel size of $k=3$.  The 2D CNN contains one convolutional layer with $k=3$ and $n_{f}=2$. We use $stride=1$ and the padding is set to half of the filter size (rounded down). The DC layer is followed by the 2D CNN layer, which forces the sampled data points in predicted k-space data to be consistent with that in the inputs.

For model training, the number of iteration step is set to $N=5$. The batch size is set to 1. All training data and test data are normalized to [0,1]. The SelfCoLearn framework with CRNN and k-t NEXT is implemented in PyTorch 1.8.1, and that with SLR-Net is implemented in Tensorflow2.2.0. The experiments are performed on a GPU server with an Nvidia Titan Xp Graphics Processing Unit (GPU, 12GB memory). The model is trained by Adam optimizer \cite{kingma2015adam} with parameters $\beta_{1}=0.5$ and $\beta_{2}=0.999$. The weight parameter $\gamma$ in co-training loss is set to 0.01. The learning rate is set to $10^{-4}$. It takes 42 hours to train SelfCoLearn with CRNN for 40 epochs and each cardiac MR data sequence takes roughly 0.5 second to get the reconstructed result.

\subsection{Comparisons to State-of-the-Art Unsupervised Methods}
To demonstrate the superiority of the proposed SelfCoLearn, we compared it with two self-supervised methods, SS-DCCNN and SS-CRNN, at different acceleration factors. It is worth noting that the state-of-the-art self-supervised method SSDU \cite{yaman2020self} was developed for static MR imaging. Literature \cite{acar2021self} adopted a similar self-supervised training manner as SSDU for dynamic MR imaging. It evaluated several backbone architectures for dynamic MR imaging including DCCNN and CRNN, whereas SSDU adopted ResNet as backbone network. We choose two self-supervised learning methods SS-DCCNN and SS-CRNN \cite{acar2021self} for comparison. In this experiment, the proposed SelfCoLearn selects the CRNN as the backbone network. For fair comparison, all methods are carefully tuned to obtain the best performance on the current dataset.

\begin{figure*}[htbp] 
\centering\includegraphics[width=\textwidth]{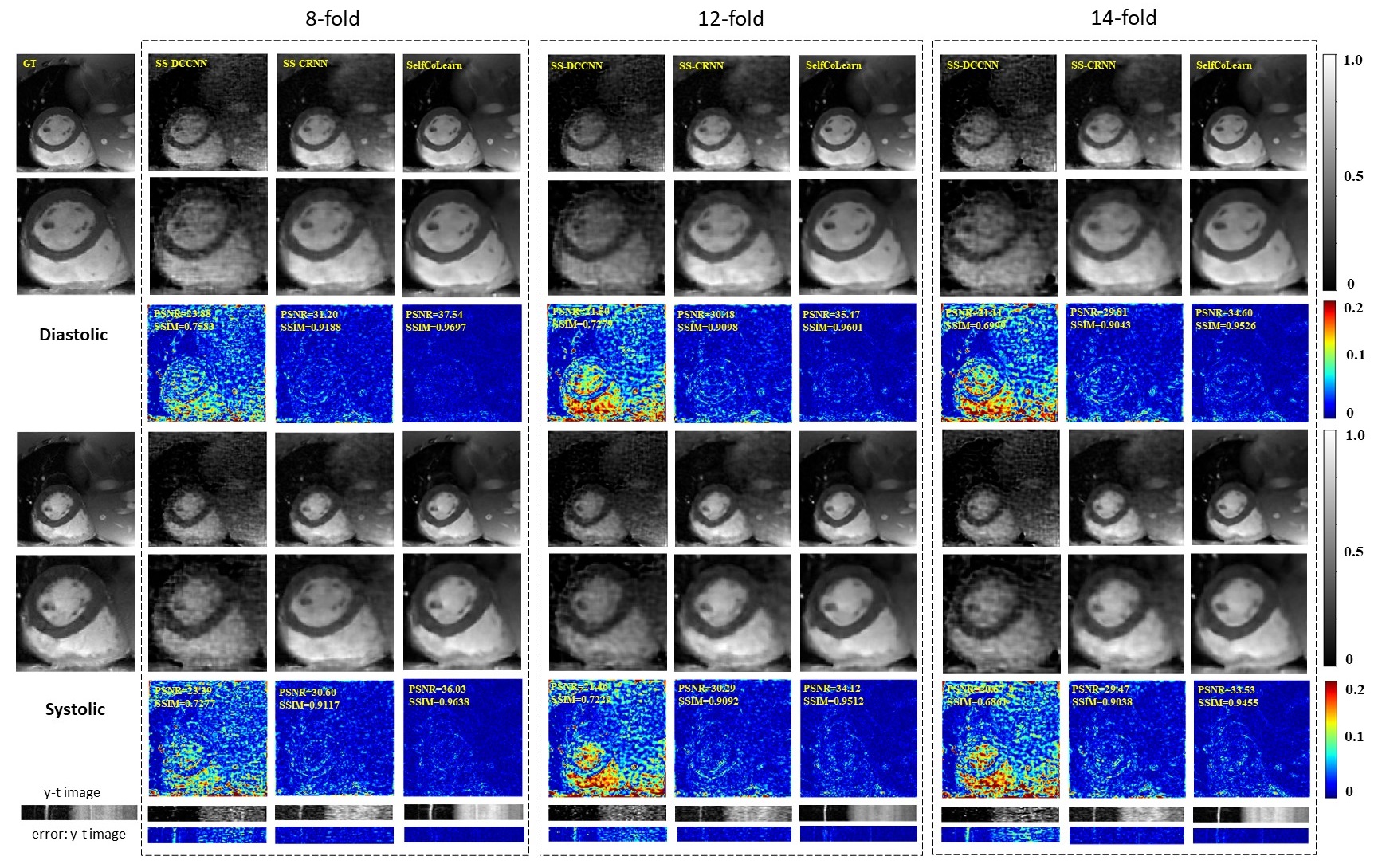} 
\caption{Reconstruction results of different methods (SS-DCCNN, SS-CRNN, and the proposed SelfCoLearn) at 8-fold acceleration, 12-fold acceleration and 14-fold acceleration. The first three rows show diastolic reconstruction results at the tenth frame. The first row shows, from left to right, fully sampled image and the reconstruction results of respective methods (display range [0, 1]). The second row shows the enlarged views of the heart regions. The third row shows the error maps of corresponding methods (display range [0, 0.2]). The following three rows show systolic reconstruction at the fifth frame. The last two rows show y-t views (extraction of the 40th slice along the y and t dimensions) and the error maps.}
\label{fig3}
\end{figure*}

Fig. \ref{fig3} plots the reconstruction results of different methods at 8-fold acceleration, 12-fold acceleration and 14-fold acceleration, respectively. The first three rows show diastolic reconstruction at the tenth frame of image sequence, and the following three rows show systolic reconstruction at the fifth frame of image sequence. The first row shows, from left to right, the fully sampled reference image and the reconstruction results of respective methods (display range [0, 1]). The second row shows the enlarged views of the heart regions. The third row shows the corresponding error maps (display range [0, 0.2]). The y-t views at $\mathbf{x}=40$ are shown in the seventh row. The corresponding error maps of y-t views are shown in the last row. From the visualization results, the proposed SelfCoLearn generates better reconstruction results than the two self-supervised methods, SS-DCCNN and SS-CRNN, at all acceleration factors. The reconstruction images of SelfCoLearn show finer structural details and more precise heart borders with fewer artifacts, while the SS-DCCNN and SS-CRNN exhibit artifacts within the heart border and chambers. The error maps of SelfCoLearn also indicate minor reconstruction errors, especially at the borders of heart chambers.

The quantitative results of these self-supervised methods are listed in Table \ref{tab1}. Similar conclusions can be drawn that the proposed SelfCoLearn achieves better quantitative performance than the two existing self-supervised learning methods. Therefore, our collaborative learning strategy can effectively capture essential and inherent representations directly from undersampled k-space data.

\begin{table*}
\begin{center}
\caption{Quantitative reconstruction results of different methods (SS-DCCNN, SS-CRNN, and the proposed SelfCoLearn) at 8-fold, 12-fold and 14-fold acceleration factors (mean±std)}
\label{tab1}
\begin{tabular}{ c || c  c | c  c  c }
AF & Methods & Training pattern & PSNR(dB) & SSIM & MSE(*e-4)\\[0.6ex]
\hline
  & SS-DCCNN & Self-supervised & 22.56±2.71 & 0.7263±0.0663 & 67.87±49.27\\[0.6ex]
8-fold   & SS-CRNN & Self-supervised & 30.81±1.77 & 0.8994±0.0288 & 9.02±3.75\\[0.6ex]

  & SelfCoLearn & Self-supervised & 37.27±2.40 & 0.9622±0.0201 & 2.17±1.22\\[0.6ex]
\hline
  & SS-DCCNN & Self-supervised & 22.17±2.76 & 0.7014±0.0658 & 74.89±54.96\\[0.6ex]

 12-fold  & SS-CRNN & Self-supervised & 30.14±1.78 & 0.8952±0.0298 & 10.54±4.40\\[0.6ex]

& SelfCoLearn & Self-supervised & 35.19±2.24 & 0.9480±0.0246 & 3.44±1.78\\[0.6ex]

\hline
  & SS-DCCNN & Self-supervised & 20.70±2.78 & 0.6667±0.0715 & 104.21±71.77\\[0.6ex]

 14-fold  & SS-CRNN & Self-supervised & 29.82±1.77 & 0.8911±0.0301 & 11.32±4.68\\[0.6ex]

& SelfCoLearn & Self-supervised & 34.38±2.23 & 0.9399±0.0273 & 4.14±2.11\\[0.6ex]
\end{tabular}
\end{center}
\end{table*}

\begin{figure}[htbp] 
\centering\includegraphics[width=\textwidth]{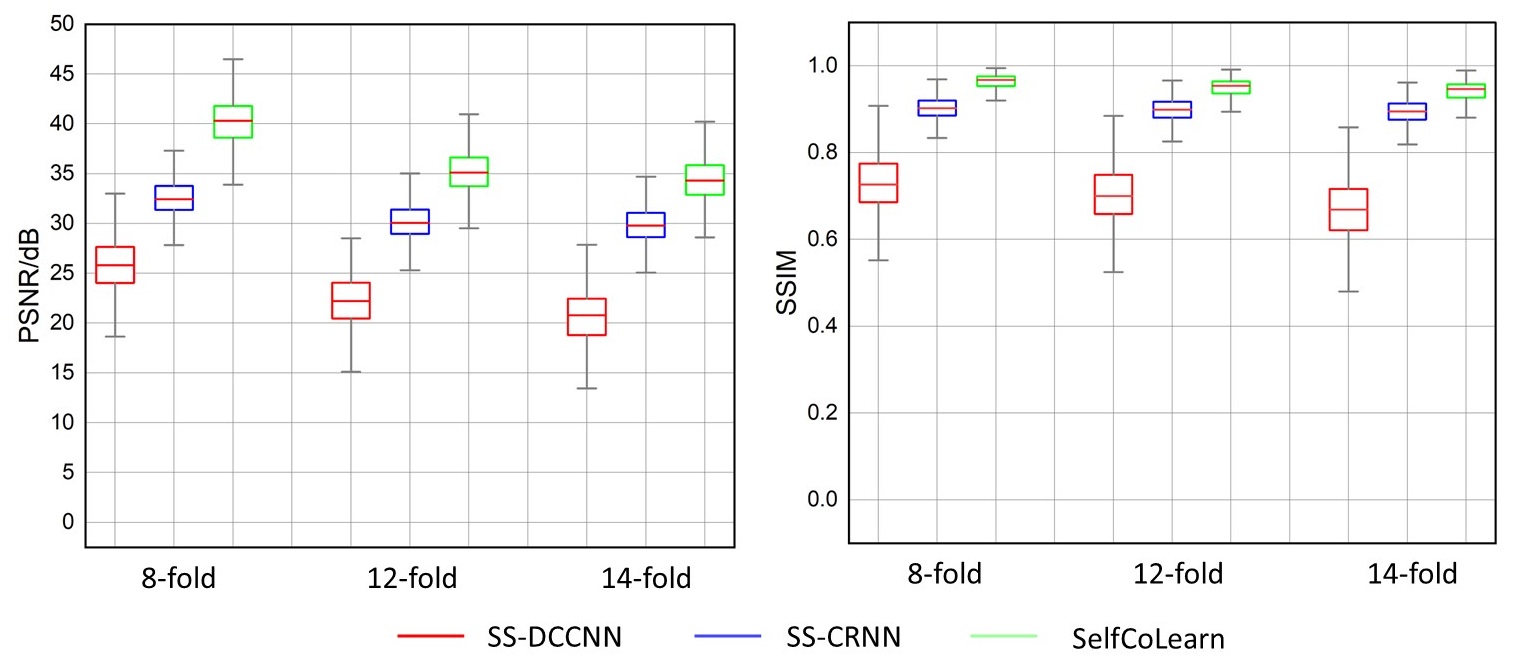} 
\caption{Box plots of different methods (SS-DCCNN, SS-CRNN and the proposed SelfCoLearn) at 8-fold, 12-fold and 14-fold accelerations, which show the median and interquartile range of PSNR and SSIM on the cardiac cine test dataset.}
\label{fig4}
\end{figure}

Fig. \ref{fig4} give the box plots displaying the median and interquartile range (25th-75th percentile) of the reconstruction results of different self-supervised methods across all test cardiac cine data at 8-fold acceleration, 12-fold acceleration and 14-fold acceleration, respectively. For all dynamic cine sequences, SelfCoLearn outperforms the two self-supervised learning methods (SS-DCCNN and SS-CRNN) at all three acceleration factors. 

\subsection{Comparisons to State-of-the-Art Supervised Methods}
\begin{figure*}[htbp] 
\centering\includegraphics[width=\textwidth]{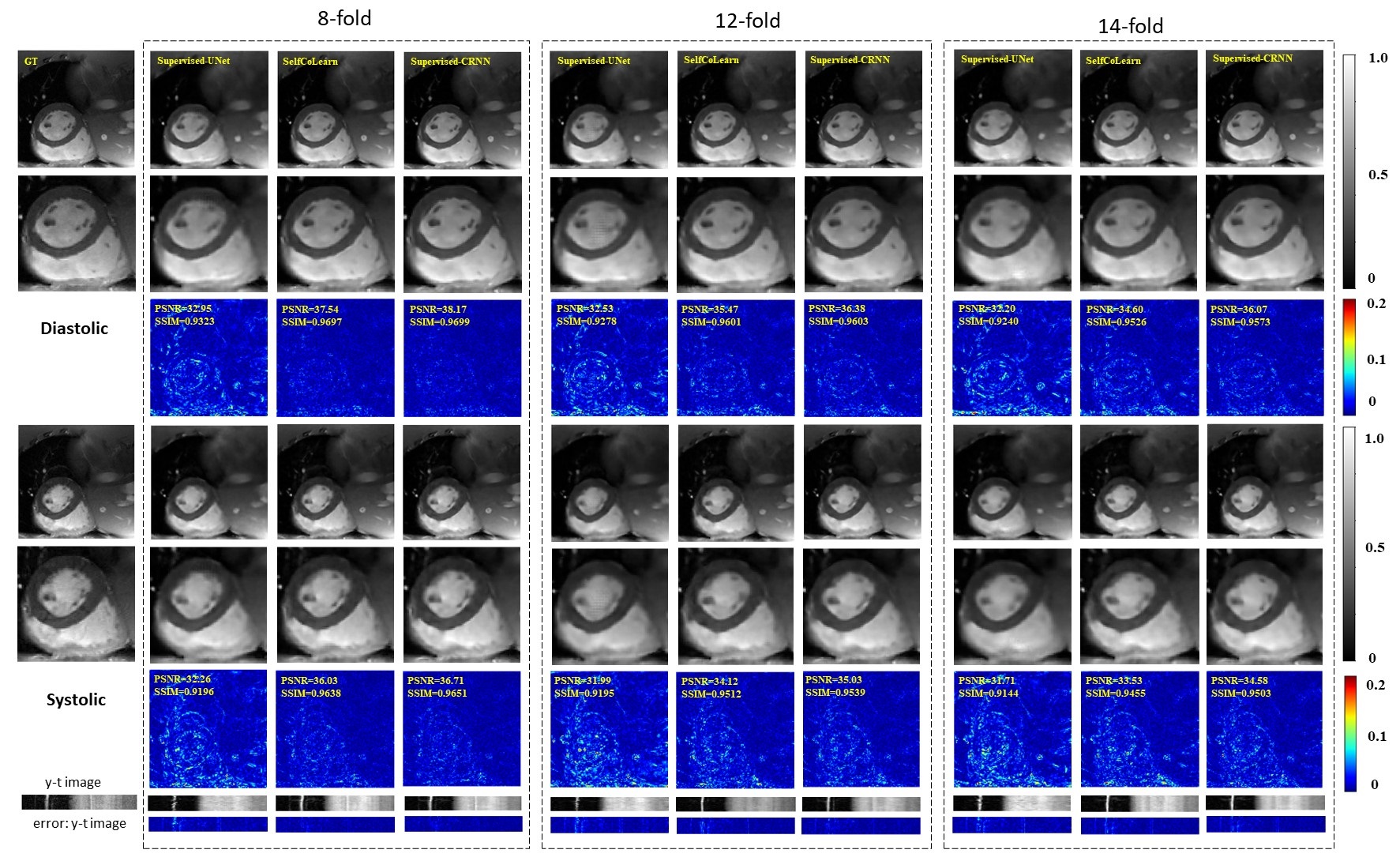} 
\caption{Reconstruction results of different methods (Supervised U-Net, Supervised CRNN and the proposed SelfCoLearn) at 8-fold acceleration, 12-fold acceleration and 14-fold acceleration. The first three rows show diastolic reconstruction results at the tenth frame. The first row shows, from left to right, fully sampled image and the reconstruction results of respective methods (display range [0, 1]). The second row shows the enlarged views of the heart regions. The third row shows the error maps of corresponding methods (display range [0, 0.2]). The following three rows show systolic reconstruction at the fifth frame. The last two rows show y-t views (extraction of the 40th slice along the y and t dimensions) and the error maps.}
\label{fig5}
\end{figure*}

We further compare our SelfCoLearn with different supervised methods, including supervised U-Net and supervised CRNN \cite{qin2018convolutional}, at different acceleration factors. Fig. \ref{fig5} plots the reconstruction results of different methods at 8-fold acceleration, 12-fold acceleration and 14-fold acceleration, respectively. From the visualization results, the proposed SelfCoLearn restores more precise anatomical details of heart regions than supervised U-Net at all acceleration factors, while supervised U-Net fails to recover some details in the heart chambers. The error maps of SelfCoLearn also indicate minor reconstruction errors than those of supervised U-Net.

In addition, SelfCoLearn generates comparable reconstruction results to those of supervised CRNN at low acceleration factors. From the enlarged heart region plots, the images reconstructed by SelfCoLearn are as clear as those generated by supervised CRNN. At higher acceleration factors, such as 14-fold acceleration, the reconstructed images of SelfCoLearn become slightly blur. Nevertheless, most of the structural details in the heart regions are still successfully restored by SelfCoLearn. Quantitative results at all acceleration factors (Table \ref{tab2}) also show the promising results of SelfCoLearn. Therefore, it can be concluded that the proposed SelfCoLearn can achieve comparable reconstruction performance with fully-supervised methods via self-supervised collaborative learning for accelerating dynamic MR imaging.

\begin{table*}
\begin{center}
\caption{Quantitative reconstruction results of different methods (Supervised U-Net, Supervised CRNN and the proposed SelfCoLearn) at 8-fold, 12-fold and 14-fold acceleration factors (mean±std)}
\label{tab2}
\begin{tabular}{ c || c  c | c  c  c }
AF & Methods & Training pattern & PSNR(dB) & SSIM & MSE(*e-4)\\[0.6ex]
\hline
& U-Net & Supervised & 32.63±1.97 & 0.9186±0.0301 & 6.06±2.88\\[0.6ex]
 8-fold & SelfCoLearn & Self-supervised & 37.27±2.40 & 0.9622±0.0201 & 2.17±1.22\\[0.6ex]
  & CRNN & Supervised & 38.09±2.52 & 0.9635±0.0204 & 1.83±1.07\\[0.6ex]
\hline
  & U-Net & Supervised & 31.96±1.88 & 0.9111±0.0317 & 6.99±3.03\\[0.6ex]
12-fold & SelfCoLearn & Self-supervised & 35.19±2.24 & 0.9480±0.0246 & 3.44±1.78\\[0.6ex]
  & CRNN & Supervised & 36.32±2.29 & 0.9513±0.0244 & 2.67±1.42\\[0.6ex]
\hline
  & U-Net & Supervised & 31.51±1.99 & 0.9045±0.0330 & 7.86±3.83\\[0.6ex]
14-fold & SelfCoLearn & Self-supervised & 34.38±2.23 & 0.9399±0.0273 & 4.14±2.11\\[0.6ex]
  & CRNN & Supervised & 35.74±2.28 & 0.9461±0.0261 & 3.05±1.59\\[0.6ex]
\end{tabular}
\end{center}
\end{table*}

\section{Discussion}
\subsection{Network Backbone Architectures}
\begin{figure}[htbp] 
\centering\includegraphics[width=\textwidth]{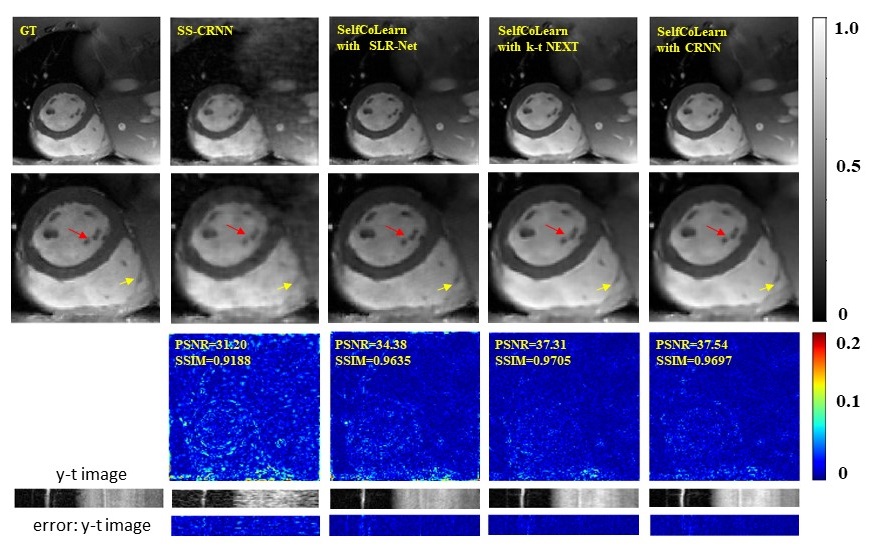}
\caption{Reconstruction results of SS-CRNN and the proposed self-supervised learning strategy with SLR-Net, k-t NEXT, and CRNN backbone networks at 8-fold acceleration. The first row shows, from left to right, fully sampled image, and the reconstruction results of SS-CRNN and the proposed self-supervised learning strategy with SLR-Net, k-t NEXT, and CRNN (display range [0, 1], 10th frame). The second row shows the enlarged views of the heart regions. The third row shows the error maps of two methods (display range [0, 0.2]). The last two rows show y-t views (extraction of the 40th slice along the y and t dimensions) and the error maps.}
\label{fig6}
\end{figure}

In this section, we explore the reconstruction results of the proposed self-supervised learning strategy with different backbone networks for dynamic MR imaging. Experiments are conducted using SLR-Net \cite{ke2021learned}, k-t NEXT \cite{qin2019k}, and CRNN \cite{qin2018convolutional} at 8-fold acceleration. The reconstruction results with different backbone networks can be found in Fig. \ref{fig6} and Table \ref{tab3}. Compared with SS-CRNN [8], the proposed self-supervised learning strategy can achieve better results regardless of the utilized backbone network. Among the three different backbone networks, SLR-Net generates worse results than k-t NEXT and CRNN. The reason for this phenomenon may be that SLR-Net needs to learn a singular value threshold, and the absence of fully sampled reference data causes the learned singular value threshold is suboptimal. However, the proposed self-supervised learning strategy with SLR-Net still obtain acceptable reconstruction results. The qualitative results in Fig. \ref{fig6}. clearly show that the proposed SelfCoLearn can better restore the structural details and achieve clearer reconstructed MR images (especially in the heart regions around the red and yellow arrows) than SS-CRNN. The quantitative results also indicate more accurate reconstructions achieved by the proposed SelfCoLearn. These results confirm that our proposed self-supervised learning framework is flexible regarding the adopted backbone network, and it can achieve promising reconstruction results with both data-driven and iterative un-rolled networks for dynamic MR imaging.

\begin{table*}
\begin{center}
\caption{Quantitative results of SS-CRNN and SelfCoLearn with different backbone networks at 8-fold acceleration(mean±std)}
\label{tab3}
\begin{tabular}{ c || c  c  c  c}
Methods & Training pattern & PSNR(dB) & SSIM & MSE(*e-4)\\[0.6ex]
\hline
SS-CRNN & Self-supervised & 30.81±1.77 & 0.8994±0.0288 & 9.02±3.75\\[0.6ex]
SelfCoLearn with SLR-Net & Self-supervised & 33.58±2.24 & 0.9495±0.0220 & 5.57±10.48\\[0.6ex]

SelfCoLearn with k-t Next & Self-supervised & 36.95±2.39 & 0.9613±0.0203 & 2.34±1.32\\[0.6ex]

SelfCoLearn with CRNN & Self-supervised & 37.27±2.40 & 0.9622±0.0201 & 2.17±1.22\\[0.6ex]

\end{tabular}
\end{center}
\end{table*}

\subsection{Co-training Loss Function}
\begin{figure}[htbp] 
\centering\includegraphics[width=\textwidth]{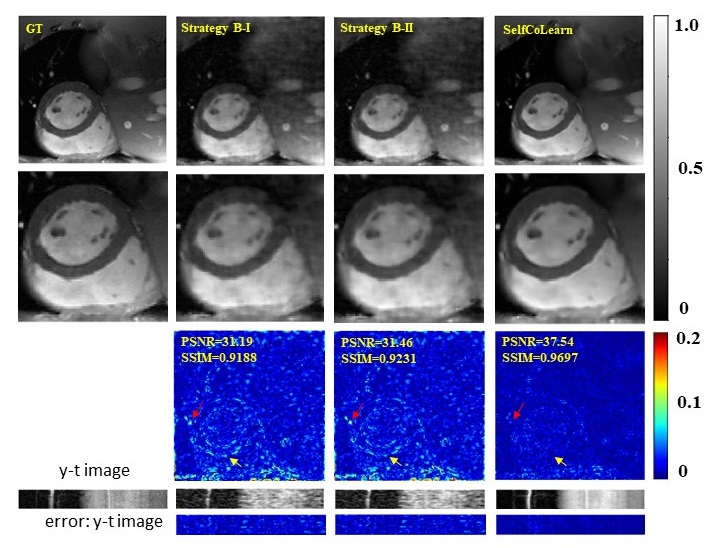}
\caption{Ablation studies utilizing different training strategies at 8-fold acceleration. The first row shows, from left to right, fully sampled image, and the reconstruction results of strategy B-I, strategy B-II and proposed SelfCoLearn (display range [0, 1], 10th frame). The second row shows the enlarged views of the heart regions. The third row shows the error maps of respective methods (display range [0, 0.2]). The last two rows show y-t views (extraction of the 40th slice along the y and t dimensions) and the error maps.}
\label{fig7}
\end{figure}

In this section, we investigate the utility of the designed the co-training loss function. The backbone network in these experiments adopts CRNN. Different training strategies at 8-fold acceleration are utilized: Strategy B-I – a single reconstruction network is trained in self-supervised manner. Only the loss function between the output $f \left( \mathbf{y}_{\Theta}^{t} \right)$ of network and $\mathbf{y}_{\Lambda}^{t}$ is used to train the network. This training strategy is similar to that of SSDU. Strategy B-II – a strategy similar to B-I but the loss function here is calculated between the output $f \left( \mathbf{y}_{\Theta}^{t} \right)$ of network and original undersampled k-space data $\mathbf{y}_{\Omega}^{t}$. SelfCoLearn – two networks are trained collaboratively with $\mathcal{L}_{U C}$ and $\mathcal{L}_{C C}$, and the two collaborative networks adopt the same backbone network as that in strategy B-I. Reconstruction images of methods utilizing the different training strategies are plotted in Fig. \ref{fig7}. Quantitative results are listed in Table \ref{tab4}. From both qualitative and quantitative results, we can observe that SelfCoLearn (training two networks collaboratively with both loss terms) achieves the best performance (especially in the heart regions around the red and yellow arrows). In particular, the contrastive consistency loss term results in a large reconstruction performance improvement. For example, PSNR is improved from 31.04dB (Strategy B-II) to 37.27 dB (SelfCoLearn).

\begin{table*}
\begin{center}
\caption{Quantitative results of reconstruction models utilizing different training strategies at 8-fold acceleration(mean±std)}
\label{tab4}
\begin{tabular}{ c || c  c  c  c c | c  c  c}
Methods & Training pattern & Single-Net & Parallel-Net & $L_{U C}$ & $L_{C C}$ & PSNR(dB) & SSIM & MSE(*e-4)\\[0.6ex]
\hline
Strategy B-I & Self-supervised & $\sqrt{ }$ & $\times$ & $\times$ & $\times$ & 30.81±1.77 & 0.8994±0.0288 & 9.02±3.75\\[0.6ex]
Strategy B-II & Self-supervised & $\sqrt{ }$ & $\times$ & $\sqrt{ }$ & $\times$ & 31.04±1.74 & 0.9045±0.0274 & 8.53±3.50\\[0.6ex]

SelfCoLearn & Self-supervised & $\times$ & $\sqrt{ }$ & $\sqrt{ }$ & $\sqrt{ }$ & 37.27±2.40 & 0.9622±0.0201 & 2.17±1.22\\[0.6ex]

\end{tabular}
\end{center}
\end{table*}

\subsection{Loss Functions}
\begin{figure}[htbp] 
\centering\includegraphics[width=\textwidth]{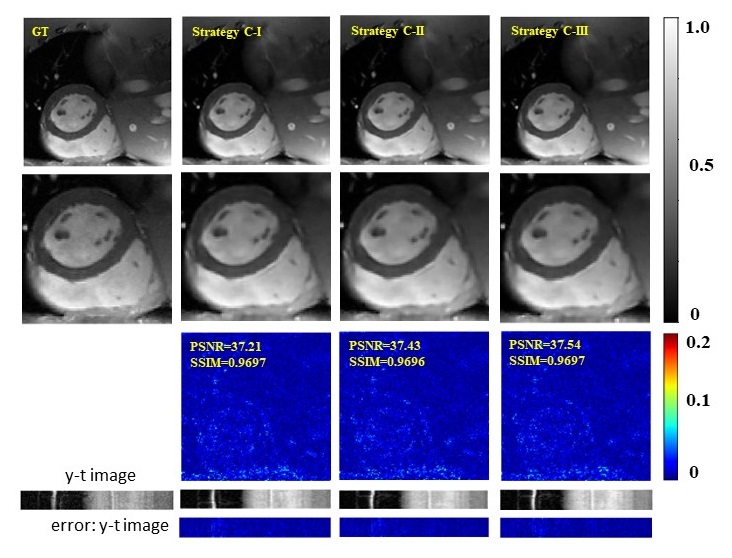}
\caption{Effects of loss functions calculated in different domains on the reconstruction results at 8-fold acceleration. The first row shows, from left to right, fully sampled image, and the reconstruction results of models utilizing Strategy C-I, C-II and C-III (display range [0, 1], 10th frame). The second row shows the enlarged views of the heart regions. The third row shows the error maps of respective strategies (display range [0, 0.2]). The last two rows show y-t views (extraction of the 40th slice along the y and t dimensions) and the error maps.}
\label{fig9}
\end{figure}

In this section, we inspect the effects of loss functions. The backbone network in these experiments adopts CRNN. Reconstruction results at 8-fold acceleration are given in Fig. \ref{fig9} and Table \ref{tab5}. Three strategies utilizing different loss function settings are investigated. In Strategy C-I, two networks are trained collaboratively with $\mathcal{L}_{U C}$ and $\mathcal{L}_{C C}$. in which $\mathcal{L}_{U C}$ is calculated in the x-t domain, and $\mathcal{L}_{C C}$ is calculated in the k-space domain. In Strategy C-II, both $\mathcal{L}_{U C}$ and $\mathcal{L}_{C C}$ are calculated in the x-t domain. In Strategy C-III, both $\mathcal{L}_{U C}$ and $\mathcal{L}_{C C}$ are calculated in the k-space domain. From both qualitative and quantitative results, we can observe that the influence of utilizing different loss function settings on the reconstruction performance is insignificant. All the other experiments in this work adopt the loss function setting of strategy C-III.

\begin{table*}
\begin{center}
\caption{Quantitative results of methods utilizing different loss function strategies at 8-fold acceleration (mean±std)}
\label{tab5}
\begin{tabular}{ c || c  c  c | c  c  c}
Methods & Training pattern & $L_{U C}$ & $L_{C C}$ & PSNR(dB) & SSIM & MSE(*e-4)\\[0.6ex]
\hline
Strategy C-I & Self-supervised & x-t domain & k-space & 37.00±2.35 & 0.9617±0.0201 & 2.30±1.29\\[0.6ex]
Strategy C-II & Self-supervised & x-t domain & x-t domain & 37.20±2.37 & 0.9617±0.0203 & 2.20±1.22\\[0.6ex]

Strategy C-III & Self-supervised & k-space & k-space & 37.27±2.40 & 0.9622±0.0201 & 2.17±1.22\\[0.6ex]

\end{tabular}
\end{center}
\end{table*}

\section{Conclusion}
In this work, we propose a self-supervised collaborative training framework to boost the image reconstruction performance for accelerating dynamic MR imaging. In particular, two independent reconstruction networks are trained collaboratively with different inputs, which are augmented from the same k-space data. To guide the two networks in capturing the detailed structural features and spatiotemporal correlations in dynamic image sequences, a co-training loss function is designed to promote the consistency between the two network predictions to provide complementary information for to-be-reconstructed dynamic MR images. The framework is flexible to be integrated with both data-driven and model-based iterative un-rolled networks. Our method has been comprehensively evaluated on a cardiac cine dataset. And comparisons to four state-of-the-art fully-supervised and self-supervised learning methods at different accelerations have been performed. SelfCoLearn achieves better results than the existing self-supervised learning methods, and it can generate comparable reconstruction performance to existing supervised learning methods. These observations indicate that SelfCoLearn possesses strong capabilities in capturing essential and inherent representations directly from the undersampled k-space data and thus enable high-quality and fast dynamic MR imaging.

\clearpage
%
%
\bibliographystyle{unsrt}
\bibliography{main}
\end{document}